\documentclass{amsart}

\usepackage[utf8]{inputenc}
\usepackage[T1]{fontenc}
\usepackage{lmodern}
\usepackage[english]{babel}

\usepackage{amsmath,amssymb,amsthm}
\usepackage{mathtools}

\usepackage{tikz}

\usepackage{hyperref}

\newcommand{\N}{\mathbb{N}} 
\newcommand{\R}{\mathbb{R}} 

\DeclareMathOperator{\Child}{Child} 
\DeclareMathOperator{\Parent}{Parent} 
\DeclareMathOperator{\Path}{Path} 
\DeclareMathOperator{\Leaves}{Leaf} 

\DeclareMathOperator{\Ex}{\mathbb{E}} 

\newcommand{\Run}{\mathcal{R}} 

\newcommand*{\defeq}{\mathrel{\vcenter{\baselineskip0.5ex \lineskiplimit0pt
                     \hbox{.}\hbox{.}}}%
                     =} 

\numberwithin{equation}{section}

\newtheorem{theorem}{Theorem}[section]
\newtheorem{lemma}[theorem]{Lemma}
\newtheorem{proposition}[theorem]{Proposition}
\newtheorem{corollary}[theorem]{Corollary}

\theoremstyle{definition}
\newtheorem{definition}[theorem]{Definition}

\theoremstyle{remark}

\title[Probability Spaces for Random Algorithms]{Probability Spaces for Random Algorithms}

\author[L. Epremidze]{Lasha Epremidze}
\email{lasha.epremidze@kiu.edu.ge}
\address{School of Mathematics, Kutaisi International University, Akhalgazrdoba Ave.~5th Lane, 4600 Kutaisi, Georgia.}

\author[G. Nadareishvili]{George Nadareishvili}
\email{giorgi.nadareishvili@kiu.edu.ge}
\address{School of Computer Science and School of Mathematics
Kutaisi International University, Akhalgazrdoba Ave.~5th Lane
4600 Kutaisi, Georgia.
}

\author[J. Oberhauser]{Jonas Oberhauser}
\email{jonas.oberhauser@huawei.com}
\address{Huawei Dresden Research Centre, Am See 3, 01067 Dresden, Germany 
}

\author[W. J. Paul]{Wolfgang J. Paul}
\email{wjp@cs.uni-sb.de}
\address{Computer Science Department,
Saarland University, Saarbr\"ucken Campus,
66123 Saarbr\"ucken, Germany.
}
\thanks{Part of this work was done while Wolfgang J. Paul was at KIU university.}


\begin{document}

\begin{abstract}
Standard analyses of expected runtimes for randomized algorithms typically bypass the explicit construction of an underlying probability space. In this paper, we provide a formal, yet intuitive tree-based definition of the probability space for the execution paths of such algorithms. Using this model, we derive the recurrence equation for the expected runtime.
\end{abstract}

\maketitle

\section{Introduction} \label{sec:intro}

The average-case analysis of randomized algorithms is almost universally conducted without the explicit construction of an underlying probability space. A prominent example dates back to one of the earliest randomized algorithms, QuickSort, introduced by Hoare~\cite{Hoa62}. He presented the randomized, single-pivot version of the algorithm and derived the famous $O(n \log n)$ bound for the expected number of comparisons $T_Q(n)$. Hoare’s proof relies on the recurrence relation
\begin{equation}\label{eq:hoare}
T_Q(n)= n-1+\frac{1}{n}\sum_{i=1}^n\big( T_Q(i-1)+T_Q(n-i)\big),
\end{equation}
which is justified by appealing to the law of conditional expectations. This is done without providing an explicit construction of the probability space governing the algorithm's executions. Although single-pivot QuickSort and its runtime analysis have long been classical textbook material, such a formal construction has, to the best of our knowledge, never appeared in standard texts or the broader research literature.

Bypassing the construction of the underlying probability space is common across the analysis of randomized algorithms, largely because the mechanics are treated as self-evident by authors. Nevertheless, for any randomized algorithm, the linearity of expectation—specifically, summing the expectations of multiple random variables as is standard practice—formally requires a common underlying probability space on which all such variables are defined. Given that probability theory is occasionally prone to unintuitive behavior, a rigorous and complete formalization is mathematically worthwhile. Ultimately, this paper's contribution is one of mathematical foundation rather than algorithmic discovery, providing an explicit probability space that formalizes the often-bypassed probabilistic mechanics of average-case algorithmic analysis.

In Section~\ref{sec:framework}, we close the aforementioned gap by constructing a general probabilistic model for randomized algorithms. Our model, which shares structural characteristics with the Galton-Watson process (see, for example~\cite{Har02}), is based on weighted random rooted trees (WRRTs). This tree-based formulation provides an intuitive, graph-theoretic picture of algorithmic execution paths. Utilizing this formalization, we  derive and prove a general recurrence equation for the expected runtime of any algorithm modeled within the framework.

To demonstrate applicability of this model, Section~\ref{sec:example_algorithms} applies the developed framework to three distinct randomized algorithms: classical QuickSelect for finding the rank of an element~\cite{Hoa61}, multi-pivot QuickSort, and Welzl's algorithm~\cite{Wel91} for computing the minimum enclosing circle from computational geometry. 

For multi-pivot QuickSort, we specifically analyze the sequential, left-pivot-first strategy for comparisons; however, we emphasize that this is merely a methodological choice, and the WRRT framework readily accommodates other implementation strategies. Such theoretical rigor remains relevant as algorithmic strategies evolve: while multi-pivot QuickSort was generally considered impractical following Hennequin's 1991 analysis~\cite{Hen91}, Vladimir Yaroslavskiy's efficient dual-pivot implementation for the Java Virtual Machine renewed interest in multi-pivot partitioning and comparison strategies (for example,~\cite{ADK15, HIKLN26, KLMA14} and~\cite{NWM16}).

Furthermore, to showcase the pedagogical value of our formalization, we provide a proof of a probabilistic identity for classical single-pivot QuickSort that is usually only intuitively justified in textbooks. Ultimately, by applying our proven recurrence equation to these diverse examples, we recover their well-known expected runtime bounds from a formalized mathematical foundation.

 {\bf Acknowledgement.} The authors thank Kurt Mehlhorn for very helpful hints and discussions.

\section{A probabilistic model} \label{sec:framework}
\subsection{Random rooted trees}
To model random algorithms  we introduce the following general framework.

Let $T$ be a \emph{rooted tree}; that is, a connected acyclic graph equipped with a distinguished vertex $O$, referred to as the root. Let $V(T)$ and $E(T)$ be the sets of vertexes and edges of $T$, respectively. The trivial case $T=O$, $V(T)=\{O\}$, $E(T)=\emptyset$ is not excluded. 

Although $T$ is defined as an undirected tree, the distinguished root $O$ induces a canonical orientation of edges away from $O$. Consequently, every vertex $v\in V(T)$ is connected to $O$ by a unique simple path \(\Path(v)\defeq (v_0,v_1,v_2,\ldots,v_{m-1},v_m)\) with $O=v_0$ and $v_m = v$.

If $v\neq O$ and $v_{m-1}$ exists in \(\Path(v)\), then the latter is called the \emph{parent} of $v$ and denoted by $\Parent(v)$. 
The set of all parents in $T$ is denoted by $\Parent(T)$.
In addition, let 
\[\Child(v)=\{u\in V(T)\mid \Parent(u)=v\}.\] We will sometimes write $\Child^T(v)$ to emphasize the tree where vertices are considered.
A vertex $v\in V(T)$, is called a \emph{leaf} if it has no children. The set of all leaves is denoted by $\Leaves(T)$. Observe that $V(T)=\Parent(T)\sqcup\Leaves(T).$

If we want to describe the path from $v_i$ to $v_j$ in \(\Path(v)\), where $0\leq i<j\leq m$, we will write \(\Path(v_i,v_j)\).

We assume that each path from the root to a leaf corresponds to a specific execution of the algorithm. Since each vertex $v \in \Leaves(T)$ uniquely determines the path $\Path(v)$ from $O$ to $v$, we encode every execution of the algorithm by its terminal vertex $v$.

The randomness on $T$ is introduced by defining the function $p\colon E(T)\to (0,1]$ that satisfies 
\begin{equation}\label{eq:P1}
\sum_{u\in\Child(v)}p(v,u)=1\quad \text{for each } v\in V(T)\setminus\Leaves(T).
\end{equation}

We assume that whenever an execution of the algorithm reaches an intermediate state $v \notin \Leaves(T)$, the subsequent state is chosen randomly from among the children of $v$ according to the corresponding transition probabilities.

The function $p$ naturally determines the map $P\colon V(T)\to(0,1]$ defined by the equations:
\begin{equation}\label{eq:defP}
P(O)=1\quad\text{and}\quad P(v)=\prod_{i=1}^{m}p(v_{i-1},v_i),
\end{equation}
where $v_i$, $i=0,1,\ldots,m$, are vertices in \(\Path(v)\). In fact, $P(v)$ denotes the probability that the algorithm reaches the state $v$ during its execution.

We call a pair $(T,p)$ a \emph{random rooted tree}.

It can be easily proved by induction that the triple $(\Leaves(T), 2^{\Leaves(T)}, P)$ is a probability space.

\begin{proposition} \label{prop:rrt}
Let $(T,p)$ be a random rooted tree and $P$ be as in \eqref{eq:defP}. Then
\begin{equation}\label{eq:P2}
    \sum_{v\in\Leaves(T)}P(v)=1.
\end{equation}
\end{proposition}

\begin{proof} Proof is by induction on the number~$n$ of vertices in $\Parent(T)$. If $n=0$, then $T=\{O\}$ and~\eqref{eq:P2} holds by definition. If $\lvert \Parent(T)\rvert =n+1$, choose~$v$ such that $\Child(v)=\{u_1,u_2,\ldots,u_m\}\subset \Leaves(T)$ and let $T'$ be the rooted tree with the same root $O$, obtained from $T$ 
by removing the vertices $u_1, u_2, \ldots, u_m$ together with the corresponding edges $(v,u_i)$, $i = 1,2,\ldots,m$.
Then $T'$ has $n$ parents since only $v$ is eliminated from $T$ as a parent, and $\sum_{u\in\Leaves(T')}P(u)=1$ by inductive assumption. Hence, by~\eqref{eq:P1} 
\begin{align*}
    \sum_{u\in\Leaves(T)}P(u) & =
    \sum_{u\in\Leaves(T)\setminus\{u_1,\ldots,u_m\}}P(u)+\sum_{i=1}^mP(u_i) \\
    & = \sum_{u\in\Leaves(T)\setminus\{u_1,\ldots,u_m\}}P(u)
    +\sum_{i=1}^mP(v)\cdot p(v,u_i) \\ 
    & =
    \sum_{u\in\Leaves(T)\setminus\{u_1,\ldots,u_m\}}P(u)
     +P(v)=\sum_{u\in\Leaves(T')}P(u)=1. \qedhere 
\end{align*}  
\end{proof}

For a rooted tree $T$ and a vertex $v\in V(T)$, let~$T_v$ denote the rooted subtree of~$T$ with root~$v$. That is, 
$T_v$ consists of $v$ together with all its descendants in~$T$, equipped with the induced edge structure. If $v=O$, then $T_v=T$, and if $v\in\Leaves(T)$, then $T_v=\{v\}$ is trivial. 

We also consider the graph $T \setminus T_v$, obtained by removing the subtree $T_v$ from $T$, however, we assume that the vertex  $v\in V(T \setminus T_v)$. Then 
$T \setminus T_v$ is again a rooted tree. If $v_1, v_2, \ldots, v_m$ are vertices of $T$ such that the subtrees 
$T_{v_1}, T_{v_2}, \ldots, T_{v_m}$ are pairwise disjoint, we naturally extend 
the above definition and consider
$T \setminus \bigcup_{i=1}^{m} T_{v_i}.$
For brevity, we denote this graph by $T \setminus \{v_1, v_2, \ldots, v_m\}.$
Observe that
\begin{equation}\label{eq:TminT}
    T \setminus \{v_1, v_2, \ldots, v_m\}
=
\bigl(T \setminus \{v_1, v_2, \ldots, v_{m-1}\}\bigr) \setminus T_{v_m}.
\end{equation}
In the notation \eqref{eq:TminT}, we always assume that $T_{v_1}, T_{v_2}, \ldots, T_{v_m}$ are pairwise disjoint. Obviously, $ T'\defeq T \setminus \{v_1, v_2, \ldots, v_m\}$ has the following property:
\begin{equation}\label{eq:S_T}
    \text{for all }v\in T', \text{ either }  v\in \Leaves(T')\text{ or } \Child^T(v)\subset V(T').
\end{equation}
See Figure~\ref{fig:T/T_v}.
\begin{figure}[htb]    
    \centering
    \begin{tikzpicture}[
      solidnode/.style={circle, draw,  minimum size=3mm, inner sep=0pt, fill=white},
      dashednode/.style={circle, draw, dashed, minimum size=3mm, inner sep=0pt},
      level 1/.style={sibling distance=1.8cm, level distance=1cm},
      level 2/.style={sibling distance=0.6cm, level distance=1cm},
      level 3/.style={sibling distance=0.8cm, level distance=1cm},
      edge from parent/.style={draw}
    ]

    \begin{scope}[xshift=0cm]
      \node[solidnode, label=above:$O$] {}
        child { node[solidnode, label={[label distance=-1mm]above left:$v$}] {}
          child { node[solidnode] {}
            child { node[solidnode] {} }
            child { node[solidnode] {} }
          }
          child { node[solidnode] {} }
          child { node[solidnode] {} }
        }
        child { node[solidnode] {}
          child { node[solidnode] {} }
          child { node[solidnode] {} }
        };
      \node[font=\large] at (0, -3.5) {(a)};
    \end{scope}

    \begin{scope}[xshift=4cm]
      \node[solidnode, label=above:$O$] {}
        child { node[solidnode, label={[label distance=-1mm]above left:$v$}] {}
          child { node[dashednode] {}
            child { node[dashednode] {} edge from parent[dashed] }
            child { node[dashednode] {} edge from parent[dashed] }
            edge from parent[dashed]
          }
          child { node[dashednode] {} edge from parent[dashed] }
          child { node[dashednode] {} edge from parent[dashed] }
        }
        child { node[solidnode] {}
          child { node[solidnode] {} }
          child { node[solidnode] {} }
        };
      \node[font=\large] at (0, -3.5) {(b)};
    \end{scope}

    \begin{scope}[xshift=8cm]
      \node[solidnode, label=above:$O$] {}
        child { node[solidnode, label={[label distance=-1mm]above left:$v$}] {}
          child { node[dashednode] {}
            child { node[dashednode] {} edge from parent[dashed] }
            child { node[dashednode] {} edge from parent[dashed] }
            edge from parent[dashed]
          }
          child { node[solidnode] {} }
          child { node[solidnode] {} }
        }
        child { node[solidnode] {}
          child { node[solidnode] {} }
          child { node[solidnode] {} }
        };
      \node[font=\large] at (0, -3.5) {(c)};
    \end{scope}

    \end{tikzpicture}%
    \caption{An illustration of removing a subtree. Panel (a) shows a rooted tree $T$. Panel (b) illustrates the tree resulting from the operation $T\setminus T_v$, where dashed lines represent the removed portion. Panel (c) does not represent a valid subtree removal and therefore does not satisfy the property~\eqref{eq:S_T}.}
    \label{fig:T/T_v}
\end{figure}

For a rooted tree $T$, let $\mathcal{S}^T$ denote the collection of all rooted subtrees $T'$ of $T$ that share the same root $O$ and satisfy the property \eqref{eq:S_T}. Clearly, every $T' \in \mathcal{S}^T$ has the form \eqref{eq:TminT} for some 
vertices $v_1, v_2, \ldots, v_m$ of $T$ and, due to properties~\eqref{eq:TminT} and~\eqref{eq:S_T}, the pair $(T',p)$ is again a random rooted tree with the same root $O$.

 Since Proposition~\ref{prop:rrt} holds for every random rooted tree, we obtain the following corollaries.

\begin{corollary}\label{cor:cor1}
    For every $T'\in S^T$, we have that  $(\Leaves(T'), 2^{\Leaves(T')}, P)$ is a probability space.
\end{corollary}
\begin{corollary}
    For every $v\in V(T)$, we have that  $(\Leaves(T_v), 2^{\Leaves(T_v)}, P_v)$ is a probability space, where $P_v$ is defined on $\Leaves(T_v)$ by the equation $P_v(u)=\prod_{i=0}^{m}p(v_{i-1},v_i),$
where  $(v =v_0,v_1,\ldots,v_m=u)$ is the path $\Path(v,u)$. That is, \(P_v(u)=P(u)/P(v).\)
Furthermore, 
\begin{equation}\label{eq:ColD}
    P\big(\Leaves(T_v)\big)=P(v).
\end{equation}
\end{corollary}

\subsection{Weighted random rooted tree}
We now introduce weights on $E(T)$ by defining a function
\[
w \colon E(T) \to \R_+.
\]
We assume that for each $(v,u) \in E(T)$, the algorithm requires $w(v,u)$ units of time to transition from $v$ to $u$. The weight function $w$ induces the so\nobreakdash-called \emph{runtime} function $\Run^T \colon V(T) \to \R_+$, defined by \(\Run^T(O)=0,\) and
\[
\Run^T(v)=\sum_{i=1}^{m}w(v_{i-1},v_i), \text{ where }  (O=v_0,v_1,\ldots,v_m=v) \text{ is the }\Path(v).
\]

A triple $(T,p,w)$ is called a {\em weighted random rooted tree} (WRRT) associated with the random algorithm. Obviously, if $(T,p,w)$ is a WRRT, then $(T',p,w)$, for each $T'\in S^T$,  and $(T_v,p,w)$, for each $v\in V(T)$, are WRRTs as well. 
It is natural to define the average runtime of the algorithm by
\begin{equation}\label{eq:EoR}
    \Ex(\Run^T)\defeq\sum_{v \in \Leaves(T)} \Run^T(v)\cdot P(v).
\end{equation}
That is, the expectation of $\Run^T$ is taken with respect to the probability space $(\Leaves(T), 2^{\Leaves(T)}, P)$.

In general, a direct application of the formula \eqref{eq:EoR} may be difficult, since the tree $T$ can grow exponentially. However, if $T$ exhibits certain self\nobreakdash-similarity patterns, a convenient recursive formula for $\Ex(\Run^T)$ can be derived as demonstrated in Section~\ref{sec:example_algorithms}. 

Note that for each $v\in V(T)$, the restriction of weight function gives the runtime function \(\Run^{T_v}\) defined on the WRRT $(T_v,p,w)$. That is, $\Run^{T_v}(v)=0,$ and for $u\in V(T_v),$
\[
 \Run^{T_v}(u)=\sum_{i=1}^{m}w(v_{i-1},v_i), \text{ where }  (v=v_0,v_1,\ldots,v_m=u) \text{ is }\Path(v,u).
\]
Note that \(\Run^{T_v}(u)=\Run^T(u)-\Run^T(v).\)

The expectation of the random variable \(\Run^{T_v}\) with respect to the probability space  $(T_v, 2^{\Leaves(T_v)}, P_v)$ is 
\[
\Ex\big(\Run^{T_v}\big)
=
\sum_{u \in \Leaves(T_v)} \Run^{T_v}(u)\cdot P_v(u).
\]

The following lemma can also be proven by the law of total expectation, however we choose a more direct approach.
\begin{lemma}
    Let $(T,p,w)$ be a weighted random rooted tree and $v\in V(T)$. Then
\begin{equation}\label{eq:5.1}
     \Ex(\Run^{T})=  \Ex(\Run^{T\setminus T_v})+P(v)
      \cdot \Ex(\Run^{T_v}).
\end{equation}
\end{lemma}
\begin{proof}
    We prove the equation \eqref{eq:5.1} by induction with respect to $n=\lvert\Parent(T_v)\rvert$. 
For $n=0$, that is, for the case where $v\in\Leaves(T)$, the relation \eqref{eq:5.1} is trivial. Suppose that the statement holds for every subtree $T_v$ with 
$\lvert \Parent(T_v)\rvert = n \ge 0$. 
Consider now a random rooted tree $T$ and a fixed vertex 
$v \in V(T)$ such that 
$\lvert \Parent(T_v)\rvert = n+1$. Choose a vertex $u \in V(T_v)$ such that 
$\Child(u) = \{u_1,u_2,\ldots,u_m\} \subset \Leaves(T_v)$, and consider the reduced tree $\widehat{T}$ obtained from $T$ by removing the vertices $u_1,u_2,\ldots,u_m$ together with the corresponding edges $(u,u_i)$, $i=1,2,\ldots,m$ (see Figure~\ref{fig:tree_diagram}). Then $\widehat{T}_v \defeq (\widehat{T})_v$ has~$n$ parents, since only~$u$ is no longer a parent, and hence, by the inductive hypothesis, we have
\begin{equation}\label{eq:5.4}
     \Ex(\Run^{\widehat{T}})=  \Ex(\Run^{T\setminus T_v})+P(v)\cdot
      \Ex(\Run^{\widehat{T}_v}),
\end{equation}
since for the middle term above we have $T \setminus T_v = \widehat{T}\setminus \widehat{T}_v $.

\begin{figure}[htpb]
    \centering
    \begin{tikzpicture}[
        basic node/.style={circle, draw,  minimum size=8mm, inner sep=0pt},
        empty node/.style={minimum size=8mm, inner sep=0pt},
        bold node/.style={circle, draw,  minimum size=8mm, inner sep=0pt},
        dashed node/.style={circle, draw, dashed, minimum size=8mm, inner sep=0pt},
        basic edge/.style={},
        bold edge/.style={},
        dashed bold edge/.style={dashed}
    ]

    \node[empty node] (L2R) at (-2, -1) {\dots};
           
    \node[bold node] (u) at (-0.5, -2) {$u$};
    \draw[bold edge] (L2R) -- (u);
    
    \node[dashed node] (u1) at (-4, -4.5) {$u_1$};
    \node[dashed node] (u2) at (-1, -4.5) {$u_2$};
    \node[dashed node] (um) at (4.5, -4.5) {$u_m$};
    
    \draw[dashed bold edge] (u) -- (u1) 
        node[midway, above left, inner sep=2pt] {$p_1$} 
        node[midway, below right, inner sep=2pt] {$w_1$};
        
    \draw[dashed bold edge] (u) -- (u2) 
        node[midway, left, inner sep=3pt] {$p_2$} 
        node[midway, right, inner sep=3pt] {$w_2$};
        
    \draw[dashed bold edge] (u) -- (um) 
        node[midway, below left, inner sep=2pt] {$p_m$} 
        node[midway, above right, inner sep=2pt] {$w_m$};
    
    \node at (0.5, -3.5) {\dots};
    
    \end{tikzpicture}
    \caption{Nodes and edges to be removed are dashed.}
    \label{fig:tree_diagram}
\end{figure}
Suppose
\[
p(u,u_i)=p_i \;\;\text{ and }\;\; w(u,u_i)=w_i,\;\;\;i=1,2,\ldots,m.
\]
Suppose also that $\Path(v,u)$ is given by $(v=v_0,v_1,\ldots,v_k=u)$ and let
\[
\pi\defeq\prod_{i=1}^{k}p(v_{i-1},v_i)=P(u)/P(v) \quad\text{and} \quad
\omega\defeq\sum_{i=1}^{k}w(v_{i-1},v_i).
\]
By definition of expectation and~\eqref{eq:P1},
\begin{multline*}
    \Ex(\Run^{T})=
    \Ex(\Run^{\widehat{T}})-P(u)\cdot \Run^T(u)+
    \sum_{i=1}^m P(u) \cdot p_i (\Run^T(u)+w_i)\\
    = \Ex(\Run^{\widehat{T}})+
    \sum_{i=1}^m P(u) \cdot p_i w_i
    =\Ex(\Run^{\widehat{T}})+
    P(v) \pi \cdot \sum_{i=1}^m p_i w_i.
\end{multline*}
Hence, by~\eqref{eq:5.4},
\begin{equation}\label{eq:4.7}
   \Ex(\Run^{T})= 
   \Ex(\Run^{T\setminus T_v})+P(v)\cdot
      \Ex(\Run^{\widehat{T}_v})
      + P(v)\pi\cdot \sum_{i=1}^m p_iw_i.
\end{equation}
Similarly,
\[
\Ex(\Run^{T_v})=
\Ex(\Run^{\widehat{T}_v})-\pi\omega+\sum_{i=1}^m \pi p_i(\omega+w_i)=
\Ex(\Run^{\widehat{T}_v})+
\pi\sum_{i=1}^m  p_iw_i.
\]
Substituting $\Ex(\Run^{\widehat{T}_v})=
\Ex(\Run^{T_v})-
\pi\sum_{i=1}^m  p_iw_i$ in~\eqref{eq:4.7}, we obtain~\eqref{eq:5.1}.
\end{proof}

The following corollary follows immediately by induction and relation~\eqref{eq:TminT}.

\begin{corollary}\label{cor:cor3}
Let $(T,p,w)$ be a WRRT and let $v_1, v_2, \dots, v_m \in V(T)$ be such that the subtrees 
$T_{v_1}, T_{v_2}, \ldots, T_{v_m}$ are pairwise disjoint. Then
\[
\Ex\bigl(\Run^{T}\bigr)
=
\Ex\bigl(\Run^{T \setminus \{v_1,v_2,\ldots,v_m\}}\bigr)
+
\sum_{i=1}^{m} P(v_i)\cdot
\Ex\bigl(\Run^{T_{v_i}}\bigr).
\]
In particular, for~$n$ children of the root $\Child(O) = \{u_1,\dots, u_n\},$ we have that
\begin{equation}\label{eq:exp_children}
\Ex\bigl(\Run^{T}\bigr)
=
\sum_{i=1}^{n} p(O, u_i)\big(w(O,u_i)
+\Ex(\Run^{T_{u_i}})\big).
\end{equation}
\end{corollary}

\section{Weighted random rooted trees for probabilistic algorithms}
\label{sec:example_algorithms}

In this section, we describe how to apply the framework of weighted random rooted trees to various randomized algorithms. We choose QuickSelect~\cite{Hoa61} for a classical application, and then sequential multi-pivot Quicksort in full generality to illustrate the process for sorting algorithms. We also describe the weighted random rooted tree for Welzl's algorithm~\cite{Wel91} to show that this model applies to algorithms from different domains.

\subsection{QuickSelect} 
We first discuss the simple, straightforward application of the framework to the classical QuickSelect algorithm~\cite{Hoa61}.

The QuickSelect algorithm takes as input a set $A = \{a_1, a_2, \dots, a_n\}$ of $n$ mutually distinct numbers and a target rank $k \in \{1, \dots, n\}$. The goal is to output the element in $A$ with rank $k$. First, we select a \emph{pivot} element $a_p \in A$ at random. We assume that the element of rank $i$ in $A$ has a probability $f(i)$ of being selected as the pivot. Next, the algorithm partitions $A$ into two subsets using $n-1$ comparisons:
\[A_< = \{a \in A \mid a < a_p\} \quad \text{and} \quad A_> = \{a \in A \mid a > a_p\}.\]

The rank of the pivot in $A$ is given by $r(a_p) = \lvert A_< \rvert + 1$. If $r(a_p) = k$, the algorithm terminates and outputs $a_p$. Otherwise, it recurses as
\[\text{QuickSelect}(A, k) = \begin{cases}
\text{QuickSelect}(A_>, k - r(a_p)) & \text{if } r(a_p) < k, \\ 
\text{QuickSelect}(A_<, k) & \text{otherwise.}
\end{cases}\] 

\begin{definition}
For $n, k \in \N_0$ with $k \leq n$, define the weighted random rooted tree $S(n,k)$ associated with $\text{QuickSelect}(A)$ for a set $A \subset \N$ with $\lvert A \rvert = n$. Assume, without loss of generality, that $A$ is indexed such that $a_i$ represents the element of rank $i$. The tree is defined as follows:
\begin{itemize}
\item $S(0,0)$ is a trivial WRRT consisting of a single node, the root $O$.
\item For $n \geq 1$, let $\Child(O) = \{a_1, \dots, a_n\}$, $p(O, a_i) = f(i)$, $w(O, a_i) = n-1$, and
\[S(n,k)_{a_i} = \begin{cases} 
S(n-i, k-i) & \text{if } i < k, \\  
O & \text{if } i = k, \\ 
S(i-1, k) & \text{if } i > k. 
\end{cases}\]
\end{itemize}
\end{definition}

See Figure~\ref{fig:S43} for a diagram of WRRT $S(4,3)$. 
\begin{figure}[htb]            
    \centering
    \begin{tikzpicture}[
        node style/.style={circle, draw,  fill=white, minimum size=3mm, inner sep=0pt},
        edge label/.style={pos=0.5, fill=white, inner sep=1pt, font=\scriptsize}, 
        ]
        \node[node style, label={above:$O$}] (O) at (0, 0.3) {};
        
        \node[node style, label={left:$a_1$}] (N1) at (-3.6, -1) {}; 
        \node[node style, label={left:$a_2$}] (N2) at (-1.2, -1) {}; 
        \node[node style, label={right:$a_3$}] (N3) at (1.2, -1) {};  
        \node[node style, label={right:$a_4$}] (N4) at (3.6, -1) {};  
        
        \draw (O) -- (N1) node[edge label, pos=0.6] {3};
        \draw (O) -- (N2) node[edge label] {3};
        \draw (O) -- (N3) node[edge label] {3};
        \draw (O) -- (N4) node[edge label, pos=0.6] {3};

        \node[node style] (N1_1) at (-4.8, -2) {}; 
        \node[node style] (N1_2) at (-3.6, -2) {}; 
        \node[node style] (N1_3) at (-2.4, -2) {}; 
        \draw (N1) -- (N1_1) node[edge label] {2};
        \draw (N1) -- (N1_2) node[edge label] {2};
        \draw (N1) -- (N1_3) node[edge label] {2};

        \node[node style] (N2_1) at (-1.2, -2) {}; 
        \node[node style] (N2_2) at (0, -2) {}; 
        \draw (N2) -- (N2_1) node[edge label] {1};
        \draw (N2) -- (N2_2) node[edge label] {1};

        \node[node style] (N4_1) at (2.4, -2) {}; 
        \node[node style] (N4_2) at (3.6, -2) {}; 
        \node[node style] (N4_3) at (4.8, -2) {}; 
        \draw (N4) -- (N4_1) node[edge label] {2};
        \draw (N4) -- (N4_2) node[edge label] {2};
        \draw (N4) -- (N4_3) node[edge label] {2};

        \node[node style] (N1_1_1) at (-5.4, -3) {}; 
        \node[node style] (N1_1_2) at (-4.2, -3) {}; 
        \draw (N1_1) -- (N1_1_1) node[edge label] {1};
        \draw (N1_1) -- (N1_1_2) node[edge label] {1};

        \node[node style] (N1_3_1) at (-3.0, -3) {}; 
        \node[node style] (N1_3_2) at (-1.8, -3) {}; 
        \draw (N1_3) -- (N1_3_1) node[edge label] {1};
        \draw (N1_3) -- (N1_3_2) node[edge label] {1};

        \node[node style] (N2_2_1) at (0, -3) {}; 
        \draw (N2_2) -- (N2_2_1) node[edge label] {0};

        \node[node style] (N4_1_1) at (1.2, -3) {}; 
        \node[node style] (N4_1_2) at (2.4, -3) {}; 
        \draw (N4_1) -- (N4_1_1) node[edge label] {1};
        \draw (N4_1) -- (N4_1_2) node[edge label] {1};

        \node[node style] (N4_2_1) at (3.6, -3) {}; 
        \draw (N4_2) -- (N4_2_1) node[edge label] {0};

        \node[node style] (N1_1_2_1) at (-4.2, -4) {}; 
        \draw (N1_1_2) -- (N1_1_2_1) node[edge label] {0};

        \node[node style] (N1_3_1_1) at (-3.0, -4) {}; 
        \draw (N1_3_1) -- (N1_3_1_1) node[edge label] {0};

        \node[node style] (N4_1_1_1) at (1.2, -4) {}; 
        \draw (N4_1_1) -- (N4_1_1_1) node[edge label] {0};
    \end{tikzpicture}%
    \caption{The weighted random rooted tree $S(4,3)$ for QuickSelect. Comparison weights $n-1$ label the edges.}
    \label{fig:S43}
\end{figure}

By Equation~\eqref{eq:exp_children} of Corollary~\ref{cor:cor3}, 
\[\Ex(\Run^{S(n,k)}) = n - 1 + \sum_{i=1}^{k-1} f(i) \Ex(\Run^{S(n-i,k-i)}) + \sum_{i=k+1}^{n} f(i) \Ex(\Run^{S(i-1,k)}).\]
For a uniform distribution $f(i) = 1/n$, we recover the classical recurrence:
\begin{align*}
T_S(n,k)&\defeq\Ex(\Run^{S(n,k)}) = n - 1 + \frac{1}{n} \Bigg(\sum_{i=1}^{k-1} \Ex(\Run^{S(n-i,k-i)}) + \sum_{i=k+1}^{n} \Ex(\Run^{S(i-1,k)})\Bigg)\\ &= n - 1 + \frac{1}{n} \Big(\sum_{i=1}^{k-1} T_S(n-i,k-i) + \sum_{i=k+1}^{n} T_S(i-1,k)\Big).
\end{align*}

\subsection{Sequential multi-pivot QuickSort}\label{sec:quicksort}
Before proceeding with our analysis of the sequential multi-pivot QuickSort algorithm, we introduce an auxiliary definition and state the associated lemma.

Given two WRRTs $T$ and $S$, we define their concatenation $T\circ S$ as the WRRT obtained by replacing each leaf of $T$ with a copy of $S$. That is, $(T\circ S)_v = S$ for each leaf $v\in \Leaves(T).$ This concatenation models an algorithm that executes the probabilistic process represented by $T$ followed immediately by the process represented by $S$ (see Figure~\ref{fig:conc}).

\begin{figure}[htb]    
    \centering
    \begin{tikzpicture}[
        vertex/.style={circle, draw,  fill=white, minimum size=3mm, inner sep=0pt},
        edge label/.style={font=\scriptsize, inner sep=1pt},
        level distance=1.5cm
        ]
        
        
        \begin{scope}[xshift=0cm, level 1/.style={sibling distance=1.8cm}]
            \node[vertex] {}
                child { node[vertex] {}
                    edge from parent 
                        node[edge label, above left, pos=0.4] {$p_1^T$} 
                        node[edge label, below right, pos=0.6] {$w_1^T$}
                }
                child { node[vertex] {}
                    edge from parent 
                        node[edge label, above right, pos=0.4] {$p_2^T$} 
                        node[edge label, below left, pos=0.6] {$w_2^T$}
                };
        \end{scope}
        
        \begin{scope}[xshift=2.97cm, level 1/.style={sibling distance=1.44cm}]
            \node[vertex] {}
                child { node[vertex] {}
                    edge from parent 
                        node[edge label, above left, pos=0.4] {$p_1^S$} 
                        node[edge label, below right, pos=0.7] {$w_1^S$}
                }
                child { node[vertex] {}
                    edge from parent 
                        node[edge label, left, pos=0.4] {$p_2^S$} 
                        node[edge label, right, pos=0.6] {$w_2^S$}
                }
                child { node[vertex] {}
                    edge from parent 
                        node[edge label, above right, pos=0.4] {$p_3^S$} 
                        node[edge label, below left, pos=0.7] {$w_3^S$}
                };
        \end{scope}
        
        \begin{scope}[xshift=7.2cm, 
            level 1/.style={sibling distance=3.42cm},
            level 2/.style={sibling distance=1.35cm}]
            
            \node[vertex] {}
                child { node[vertex] {}
                    child { node[vertex] {}
                        edge from parent 
                            node[edge label, above left, pos=0.4] {$p_1^S$} 
                            node[edge label, below right, pos=0.7] {$w_1^S$}
                    }
                    child { node[vertex] {}
                        edge from parent 
                            node[edge label, left, pos=0.4] {$p_2^S$} 
                            node[edge label, right, pos=0.6] {$w_2^S$}
                    }
                    child { node[vertex] {}
                        edge from parent 
                            node[edge label, above right, pos=0.4] {$p_3^S$} 
                            node[edge label, below left, pos=0.7] {$w_3^S$}
                    }
                    edge from parent 
                        node[edge label, above left, pos=0.4] {$p_1^T$} 
                        node[edge label, below right, pos=0.6] {$w_1^T$}
                }
                child { node[vertex] {}
                    child { node[vertex] {}
                        edge from parent 
                            node[edge label, above left, pos=0.4] {$p_1^S$} 
                            node[edge label, below right, pos=0.7] {$w_1^S$}
                    }
                    child { node[vertex] {}
                        edge from parent 
                            node[edge label, left, pos=0.4] {$p_2^S$} 
                            node[edge label, right, pos=0.6] {$w_2^S$}
                    }
                    child { node[vertex] {}
                        edge from parent 
                            node[edge label, above right, pos=0.4] {$p_3^S$} 
                            node[edge label, below left, pos=0.7] {$w_3^S$}
                    }
                    edge from parent 
                        node[edge label, above right, pos=0.4] {$p_2^T$} 
                        node[edge label, below left, pos=0.6] {$w_2^T$}
                };
        \end{scope}
        
        \node at (1.35cm, -3.8cm) {(a)};
        \node at (7.2cm, -3.8cm) {(b)};
        
    \end{tikzpicture}
    \caption{An example of weighted random rooted trees (a) and their concatenation (b).}
    \label{fig:conc}
\end{figure}
Corollary~\ref{cor:cor3} directly implies the following
\begin{corollary}\label{cor:conc} If $T$ and $S$ are weighted random rooted trees, then
\[\Ex(\Run^{T\circ S}) = \Ex(\Run^T)+\Ex(\Run^S). \]
\end{corollary}
\begin{proof}
By Corollary~\ref{cor:cor3},
\begin{align*} 
\Ex(\Run^{T\circ S}) 
&= \Ex(\Run^T) + \sum_{v \in \Leaves(T)}P^T(v)\Ex(\Run^S)= \Ex(\Run^T) + \Ex(\Run^S) \sum_{v \in \Leaves(T)}P^T(v)\\
&= \Ex(\Run^T) + \Ex(\Run^S).\qedhere
\end{align*}
\end{proof}

A brief specification of the general \emph{sequential} $k$-pivot QuickSort algorithm is as follows. Let the input be a set $A=\{a_1, a_2, \dots, a_n\}$ consisting of mutually distinct elements. If $n \leq k$, all elements are designated as \emph{pivots} and sorted directly. If $n > k$, we select $k$ pivots from $A$. Once chosen, the pivots are sorted such that 
\[ a_{p_1} < a_{p_2} < \dots < a_{p_k}. \]
In general, the set of pivots is chosen with a certain probability $f(\{a_{p_1}, \dots, a_{p_k}\})$. The strategy for sorting the $m \leq k$ pivot elements is fixed. 

We define the partitioned subsets as:
\[ A_1 = \{a \in A \mid a < a_{p_1}\}, \quad A_{k+1} = \{a \in A \mid a > a_{p_k}\}, \]
\[ A_i = \{a \in A \mid a_{p_{i-1}} < a < a_{p_i}\} \quad \text{for } i = 2, \dots, k. \]

The sorted result is then constructed recursively as:
\begin{multline}\label{eq:quicksort}
k\text{-QuickSort}(A) = k\text{-QuickSort}(A_1) \circ a_{p_1} \\
\circ k\text{-QuickSort}(A_2) \circ a_{p_2} \circ \dots \circ a_{p_k} \circ k\text{-QuickSort}(A_{k+1}). 
\end{multline}
To determine the rank of the pivot $a_{p_{i+1}}$, the sequential multi-pivot QuickSort compares $a_{p_{i+1}}$ against all $n-k-\sum_{m=1}^{i}\lvert A_m\rvert$ non-pivot elements that are strictly greater than $a_{p_i}$. Note that this differs from the partitioning strategy employed by Yaroslavskiy's dual-pivot QuickSort in the Java Virtual Machine, falling instead within the broader framework of comparison strategies investigated in~\cite{ADK15}.

Now we are ready to define a weighted random rooted tree corresponding to the multi-pivot QuickSort algorithm. 

\begin{definition}
For any $k, n \in \N_0$, we define the weighted random rooted tree $Q(n,k)$ associated with $k\text{-QuickSort}(A)$ for a set $A\subset \N$ with $\lvert A \rvert = n$ as follows. 
\begin{itemize}
\item For $n \leq k$, $Q(n,k)$ is the trivial WRRT consisting of a single root node $O$ and no edges. For $n > k$, let $\Child(O) \defeq \binom{A}{k}$, the set of all $k$-element subsets of $A$:
\[\binom{A}{k} = \big\{ \{x_{i_1}, x_{i_2}, \dots, x_{i_k}\} \subset A \mid i_j \text{ is the rank of } x_{i_j} \text{ in } A \text{ and } i_1<\dots <i_k \big\}.\] 
Note that in this notation, $\lvert A_1 \rvert = i_1-1$, $\lvert A_2 \rvert = i_2-i_1-1$, \dots, $\lvert A_k \rvert = i_k-i_{k-1}-1$, and $\lvert A_{k+1} \rvert = n-i_k$.

\item $p(O, \{x_{i_1}, x_{i_2}, \dots, x_{i_k}\}) \defeq f(\{x_{i_1}, x_{i_2}, \dots, x_{i_k}\})$ for a probability mass function 
\[f\colon \binom{A}{k} \to (0,1].\] 
If every $k$-element subset is equally likely to be chosen as the pivot set, then 
\[f(\{x_{i_1}, x_{i_2}, \dots, x_{i_k}\}) = \frac{1}{\binom{n}{k}}.\]

\item Let $r_m$ denote the runtime of the fixed sorting strategy for $m \leq k$ pivot elements; assume $r_m = 0$ for $m > k$. The edge weights are then given by
\begin{align*}
w(O, \{x_{i_1}, \dots, x_{i_k}\}) &= r_k + n - k + \sum_{j=1}^{k-1}(n - k - i_j + j) + \sum_{j=1}^{k+1} r_{\lvert A_j \rvert} \\
&= kn - \binom{k+1}{2} - \sum_{j=1}^{k-1} i_j + r_k + \sum_{j=1}^{k+1} r_{\lvert A_j \rvert},
\end{align*} 
where $r_k$ is the runtime for sorting the $k$ pivots, the $n-k$ term represents the number of comparisons necessary to determine the rank~$i_1$ of $x_{i_1}$ in~$A$, and $n-k-\sum_{m=1}^j \lvert A_m \rvert = n - k - i_j + j$ is the number of comparisons required to determine~$i_{j+1}$. The sum $\sum r_{\lvert A_j \rvert}$ represents the absorbed runtime cost from sorting the pivots in any resulting partitioned subsets $A_j$ of size less than or equal to~$k$. 

\item The subtrees rooted at the children of the root are given by
\[
Q(n,k)_{\{x_{i_1}, \dots, x_{i_k}\}} = 
Q(\lvert A_1 \rvert, k) \circ Q(\lvert A_2 \rvert, k) \circ \dots \circ Q(\lvert A_k \rvert, k) 
\circ Q(\lvert A_{k+1} \rvert, k).
\]
\end{itemize}
\end{definition}

\begin{figure}[htb]            
    \centering
    \begin{tikzpicture}[
        node style/.style={circle, draw,  fill=white, minimum size=3mm, inner sep=0pt},
        edge label/.style={pos=0.6, fill=white, inner sep=1pt, font=\scriptsize}
        ]
        
        \node[node style, label=above:$O$] (O) at (0, 0) {};
        
        
        \node[node style] (N12) at (-4.68, -2.5) {};
        \node[font=\scriptsize, above left=0pt] at (N12) {\{1,2\}};
        
        \node[node style] (N15) at (-1.08, -2.5) {};
        \node[font=\scriptsize, above left=0pt] at (N15) {\{1,5\}};
        
        \node[node style] (N45) at (4.68, -2.5) {};
        \node[font=\scriptsize, above right=0pt] at (N45) {\{4,5\}};

        \node[node style] (N13) at (-3.24, -2.5) {}; \node[font=\scriptsize, below=3pt] at (N13) {\{1,3\}};
        \node[node style] (N14) at (-2.52, -2.5) {}; \node[font=\scriptsize, below=3pt] at (N14) {\{1,4\}};
        \node[node style] (N23) at (0.36, -2.5) {};  \node[font=\scriptsize, below=3pt] at (N23) {\{2,3\}};
        \node[node style] (N24) at (1.08, -2.5) {};  \node[font=\scriptsize, below=3pt] at (N24) {\{2,4\}};
        \node[node style] (N25) at (1.8, -2.5) {};   \node[font=\scriptsize, below=3pt] at (N25) {\{2,5\}};
        \node[node style] (N34) at (2.52, -2.5) {};  \node[font=\scriptsize, below=3pt] at (N34) {\{3,4\}};
        \node[node style] (N35) at (3.24, -2.5) {};  \node[font=\scriptsize, below=3pt] at (N35) {\{3,5\}};
        
        \draw (O) -- (N12) node[edge label, pos=0.65] {7};
        \draw (O) -- (N13) node[edge label] {8};
        \draw (O) -- (N14) node[edge label] {8};
        \draw (O) -- (N15) node[edge label, pos=0.65] {7};
        \draw (O) -- (N23) node[edge label] {7};
        \draw (O) -- (N24) node[edge label] {6};
        \draw (O) -- (N25) node[edge label] {7};
        \draw (O) -- (N34) node[edge label] {6};
        \draw (O) -- (N35) node[edge label] {6};
        \draw (O) -- (N45) node[edge label, pos=0.65] {4};

        
        \node[node style] (N12_1) at (-5.4, -5) {}; \node[font=\scriptsize, below=3pt] at (N12_1) {\{3,4\}};
        \node[node style] (N12_2) at (-4.68, -5) {}; \node[font=\scriptsize, below=3pt] at (N12_2) {\{3,5\}};
        \node[node style] (N12_3) at (-3.96, -5) {}; \node[font=\scriptsize, below=3pt] at (N12_3) {\{4,5\}};
        \draw (N12) -- (N12_1) node[edge label] {3};
        \draw (N12) -- (N12_2) node[edge label] {3};
        \draw (N12) -- (N12_3) node[edge label] {2};

        \node[node style] (N15_1) at (-1.8, -5) {}; \node[font=\scriptsize, below=3pt] at (N15_1) {\{2,3\}};
        \node[node style] (N15_2) at (-1.08, -5) {}; \node[font=\scriptsize, below=3pt] at (N15_2) {\{2,4\}};
        \node[node style] (N15_3) at (-0.36, -5) {}; \node[font=\scriptsize, below=3pt] at (N15_3) {\{3,4\}};
        \draw (N15) -- (N15_1) node[edge label] {3};
        \draw (N15) -- (N15_2) node[edge label] {3};
        \draw (N15) -- (N15_3) node[edge label] {2};

        \node[node style] (N45_1) at (3.96, -5) {}; \node[font=\scriptsize, below=3pt] at (N45_1) {\{1,2\}};
        \node[node style] (N45_2) at (4.68, -5) {}; \node[font=\scriptsize, below=3pt] at (N45_2) {\{1,3\}};
        \node[node style] (N45_3) at (5.4, -5) {}; \node[font=\scriptsize, below=3pt] at (N45_3) {\{2,3\}};
        \draw (N45) -- (N45_1) node[edge label] {3};
        \draw (N45) -- (N45_2) node[edge label] {3};
        \draw (N45) -- (N45_3) node[edge label] {2};

    \end{tikzpicture}
    \caption{The weighted random rooted tree $Q(5,2)$ evaluating with base costs $r_2=1, r_1=0$. Because $n=5$ and $k=2$, the largest possible partitioned subset has size 3. Thus, 7 of the $\binom{5}{2}=10$ root branches immediately satisfy $\lvert A_j \rvert \le 2$ and resolve into trivial base cases. The remaining 3 branches (pivots $\{1,2\}$, $\{1,5\}$, and $\{4,5\}$) each leave exactly one partition of size 3, triggering a recursive $Q(3,2)$ subtree. Each $Q(3,2)$ subtree branches into $\binom{3}{2}=3$ final leaves which absorb the remaining sorting costs locally.}
    \label{fig:Q52}
\end{figure}

Equation~\eqref{eq:exp_children} of Corollary~\ref{cor:cor3} and Corollary~\ref{cor:conc} yield the recursive equation
\begin{multline*}
\Ex\bigl(\Run^{Q(n,k)}\bigr)
=
\sum_{\{x_{i_1}, \dots, x_{i_k}\}\in\binom{A}{k}} f(\{x_{i_1}, \dots, x_{i_k}\})\Biggl(kn - \binom{k+1}{2} \\ 
- \sum_{j=1}^{k-1} i_j + r_k + \sum_{j=1}^{k+1} \Bigl(r_{\lvert A_j \rvert}
+\Ex\bigl(\Run^{Q(\lvert A_j \rvert, k)}\bigr)\Bigr)\Biggr)
\end{multline*}
for $\lvert A \rvert =n$. Note that for $k=1$ and the uniform probability mass function $f = 1/n$, we recover the classical equation~\eqref{eq:hoare} by Hoare:
\begin{align*}
T_{Q}(n) &\defeq \Ex\bigl(\Run^{Q(n,1)}\bigr) = n-1+\frac{1}{n}\sum_{i=1}^n\Bigl( \Ex\bigl(\Run^{Q(i-1,1)}\bigr)+\Ex\bigl(\Run^{Q(n-i,1)}\bigr)\Bigr)\\ &= n-1+\frac{1}{n}\sum_{i=1}^n\bigl( T_{Q}(i-1)+T_{Q}(n-i)\bigr). 
\end{align*}

\subsubsection{Probability of direct comparison between two elements}

An intuitive justification is provided in several textbooks~\cite{MU05, MR95, SR14} for the formula
\begin{equation}\label{eq:PAij}
P(A_{i,j}) = \frac{2}{j-i+1}, \qquad 1 \le i < j \le n,
\end{equation}
where $A_{i,j}$ is the event that $a_i$ and $a_j$ are compared to each other during the execution of the classical single-pivot QuickSort algorithm under uniform pivot selection. 
In particular, this probability does not depend on the input size~$n$. In this section, we derive \eqref{eq:PAij} relying on the WRRT construction of QuickSort introduced in Section~\ref{sec:quicksort}. 
To this end, we will utilize the following simple lemma, whose proof is straightforward.

\begin{lemma} \label{lem:partition}
Let $(\Omega,\mathcal{F},\mathbb{P})$ be a probability space and let $E \in \mathcal{F}$ be an event. 
Let \(E = \bigcup_{n=1}^{\infty} E_n\) and \(E^c = \bigcup_{n=1}^{\infty} E'_n\)
be partitions of $E$ and its complement $E^c = \Omega \setminus E$ into pairwise disjoint measurable sets, respectively. If 
\(\mathbb{P}(E'_n)=c\cdot\mathbb{P}(E_n)\) for all $n \ge 1$ and some constant $c \ge 0$, then
\[
\mathbb{P}(E) = \frac{1}{1+c}.
\]
\end{lemma}

We proceed with the proof of~\eqref{eq:PAij} by considering the WRRT $Q=Q(n,1)$ and its corresponding probability space $(\Leaves(Q), 2^{\Leaves(Q)}, P)$. We identify every vertex of $Q$ with a partition of the set $A$ into subsets determined by the sequence of pivot choices made prior to reaching that node, where the previously chosen pivots themselves form singleton subsets. We assume that the elements within these partitioned subsets preserve their natural order. 

Let $V(i,j)$ be the set of vertices in~$Q$ representing states where $a_i$ and~$a_j$ still belong to the same unsorted subset. For any $v \in V(i,j)$, let $u_k$ denote the child of~$v$ produced by selecting the pivot~$a_k$ (where $i \le k \le j$). We then have
\[
\Leaves(Q_{u_i}) \cup \Leaves(Q_{u_j}) \subset A_{i,j},
\qquad
\bigcup_{i<k<j} \Leaves(Q_{u_k}) \subset A_{i,j}^c.
\]
That is, once the state~$v$ is reached, $a_i$ and~$a_j$ will only be compared to each other if we select either of them as the pivot at~$v$. Furthermore, the unions
\[
\bigcup_{v \in V(i,j)} \Bigl( \Leaves(Q_{u_i}) \cup \Leaves(Q_{u_j}) \Bigr)
\quad \text{and} \quad
\bigcup_{v \in V(i,j)} \bigcup_{i<k<j} \Leaves(Q_{u_k})
\]
form partitions of $A_{i,j}$ and $A_{i,j}^c$ into pairwise disjoint sets, respectively. 

Because the pivot is selected uniformly at random at state~$v$, the algorithm transitions to any child $u_k$ with identical probability. Consequently, by~\eqref{eq:ColD}, the sets of leaves reachable through each of these children carry the same total probability mass. Thus, 
\begin{align*}
P\Bigl( \bigcup_{i<k<j} \Leaves(Q_{u_k}) \Bigr)
&= \sum_{k=i+1}^{j-1} P\bigl(\Leaves(Q_{u_k})\bigr) \\
&= \frac{j-i-1}{2} \Bigl( P\bigl(\Leaves(Q_{u_i})\bigr) + P\bigl(\Leaves(Q_{u_j})\bigr) \Bigr) \\
&= \frac{j-i-1}{2}\, P\Bigl( \Leaves(Q_{u_i}) \cup \Leaves(Q_{u_j}) \Bigr).
\end{align*}
As this proportional relationship holds for every individual vertex $v \in V(i,j)$, we can apply Lemma~\ref{lem:partition} with $c = \frac{j-i-1}{2}$ to obtain
\[
P(A_{i,j}) = \frac{1}{1 + \frac{j-i-1}{2}} = \frac{2}{j-i+1}.
\]

\subsection{Welzl's minimum enclosing circle algorithm}
To demonstrate that the WRRT model applies to randomized geometric algorithms, we formalize the expected runtime of Emo Welzl's algorithm for finding the minimum enclosing circle (MEC) of a set of points in the plane~\cite{Wel91}.

Let $P$ be a set of $n$ points in $\R^2$, and let $R$ be a set of points known to lie on the boundary of the MEC (initially $R = \emptyset$). Welzl's algorithm, denoted $\text{MinDisk}(P, R)$, proceeds as follows:
If $P$ is empty or $\lvert R \rvert = 3$, it returns the trivial MEC of~$R$ in~$O(1)$ time. 
Otherwise, it selects a point $x \in P$ uniformly at random and recursively computes $D = \text{MinDisk}(P \setminus \{x\}, R)$. 
It then checks if $x \in D$. If it is, $D$ is the correct MEC. If $x \notin D$, then $x$ must lie on the boundary of the new MEC, and the algorithm makes a second recursive call $\text{MinDisk}(P \setminus \{x\}, R \cup \{x\})$.

Because the MEC of any set of points is uniquely defined, the condition $x \notin \text{MinDisk}(P \setminus \{x\}, R)$ is a deterministic geometric property of the sets $P$ and $R$ and the chosen point $x$, entirely independent of the random choices made during the first recursive call. Let $\chi(x, P, R) \in \{0,1\}$ be the indicator function of this event. 

Let $C$ denote a trivial WRRT consisting of a root, a single child, and one edge of weight $c$, representing the $O(1)$ time required to check if $x \in D$. We construct the WRRT for Welzl's algorithm as follows.

\begin{definition}
For a set of points $P \subset \R^2$ with $\lvert P \rvert = n$ and a boundary set $R$ with $\lvert R \rvert \leq 3$, we define the weighted random rooted tree $W(P,R)$ as follows:
\begin{itemize}
    \item If $n = 0$ or $\lvert R \rvert = 3$, $W(P,R)$ is a trivial WRRT consisting of a single root node $O$ and no edges. If $n > 0$ and $\lvert R \rvert < 3$, the root $O$ has children $\Child(O) = P$.
    \item For each $x \in P$, the transition probability is uniform $p(O, x) = 1/n$.
    \item The edges from the root carry no computational weight $w(O, x) = 0$.
    \item The subtree rooted at the child corresponding to the choice $x \in P$ is constructed via concatenation:
    \[
    W(P,R)_x = 
    \begin{cases}
    W(P \setminus \{x\}, R) \circ C & \text{if } \chi(x, P, R) = 0, \\
    W(P \setminus \{x\}, R) \circ C \circ W(P \setminus \{x\}, R \cup \{x\}) & \text{if } \chi(x, P, R) = 1.
    \end{cases}
    \]
\end{itemize}
\end{definition}

Applying Corollary~\ref{cor:cor3} and Corollary~\ref{cor:conc} to this WRRT immediately yields the expected runtime. Let $T_W(n, r) \defeq \Ex(\Run^{W(P,R)})$ where $\lvert P \rvert = n$ and $\lvert R \rvert = r$. Then,
\begin{align*}
T_W(n, r) &= \sum_{x \in P} \frac{1}{n} \Ex(\Run^{W(P,R)_x}) \\
&= \frac{1}{n} \sum_{x \in P} \Bigl( \Ex\bigl(\Run^{W(P \setminus \{x\}, R)}\bigr) + c + \chi(x, P, R) \Ex\bigl(\Run^{W(P \setminus \{x\}, R \cup \{x\})}\bigr) \Bigr) \\
&= T_W(n-1, r) + c + \frac{1}{n} \sum_{x \in P} \chi(x, P, R) T_W(n-1, r+1).
\end{align*}
A fundamental geometric property of the MEC is that at most 3 points in $P$ can strictly define the circle, meaning $\sum_{x \in P} \chi(x, P, R) \leq 3 - r \leq 3$. This gives the classical recurrence
\[
T_W(n, r) \leq T_W(n-1, r) + c + \frac{3-r}{n} T_W(n-1, r+1),
\]
which evaluates to the well-known $O(n)$ expected runtime.

\bibliographystyle{plain}  
\bibliography{bibl}

\end{document}